\documentclass[12pt]{iopart}
\usepackage{epsfig}
\begin{document}

\title{Conical Emission in Heavy Ion Collisions}

\author{Jason Glyndwr Ulery}

\address{
Institut f\"{u}r Kernphysik	\hspace{40pt}	Purdue University\\
Max-von-Laue-Stra\ss e 1	\hspace{40pt}	525 Northwestern Ave.\\
Frankfurt am Main, Germany	\hspace{12pt}	Lafayette, IN USA}
\ead{ulery@ikf.uni-frankfurt.de}
\begin{abstract}
A broadened or double humped away-side structure was observed in 2-particle azimuthal jet-like correlations at RHIC and SPS.  This modification can be explained by conical emission, from either Mach-cone shock waves or \v{C}erenkov gluon radiation, and by other physics mechanisms, such as large angle gluon radiation, jets deflected by radial flow and path-length dependent energy loss.  Three-particle jet-like correlations are studied for their power to distinguish conical emission from other mechanisms.  This article discusses Mach-cone shock waves, \v{C}erenkov gluon radiation and the experimental evidence for conical emission from RHIC and SPS.
\end{abstract}


\section{Introduction}

Mach-cones in cold nuclear matter were suggested as a possibility in heavy ion collisions over 30 years ago\cite{mach0}, but have not been observed.  In recent high energy heavy ion data, where a hot dense medium is created, a double peaked away-side structure was observed in 2-particle jet-like correlations\cite{star2p,phenix}.  This is in contrast to a single away-side peak seen in $pp$ and d+Au collisions.  Many physics mechanisms have been suggested to explain this modification of the 2-particle correlation.  These include conical emission from Mach-cone shock waves\cite{mach0,mach1,mach2,mach25,mach3,mach4,mach5,mach6} or \v{C}erenkov gluon radiation\cite{cerenkov0, cerenkov05,cerenkov1,cerenkov2}, and large angle gluon radiation\cite{gluon1,gluon2}, path-length dependent energy loss\cite{path}, and jets deflected by radial flow\cite{deflected}.  There has been a great deal of recent theoretical and experimental work to distinguish the true source of the modification.  The nature of this modification can shed light on the nature of the medium created in heavy-ion collisions.  

\section{Mach-Cone Shock Waves}
\begin{figure}[htpb]
\centering
\begin{minipage}[t]{0.33\textwidth}
\includegraphics[width=0.99\textwidth]{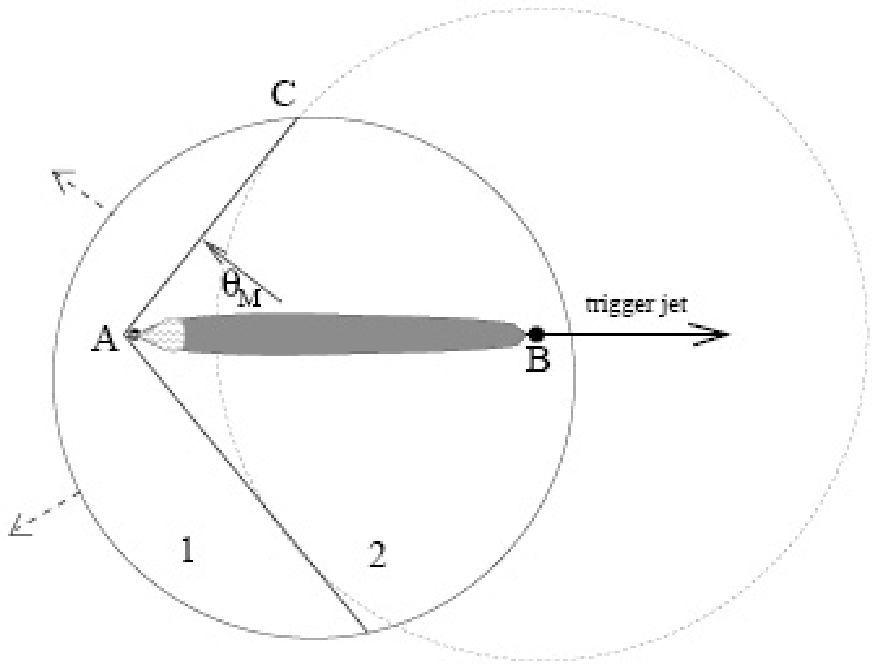}
\end{minipage}
\begin{minipage}[t]{0.35\textwidth}
\includegraphics[width=0.99\textwidth]{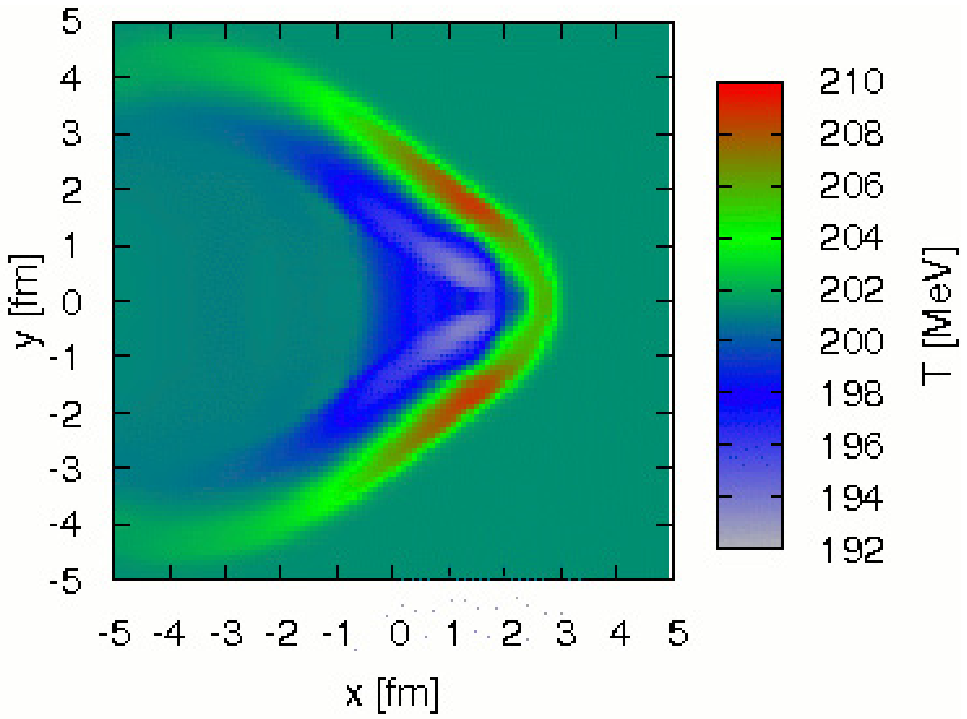}
\end{minipage}
\caption{Left:  Diagram of the away-side jet creating a Mach-cone\cite{mach2}.  Right:  Shock wave from hydrodynamic calculation\cite{betz}.}
\label{fig:hydro}
\end{figure}

If a particle exceeds the speed of sound in a medium, a Mach-cone can be formed.  By measuring the angle of a Mach-cone the speed of sound of the medium, which provides information about the equation of state, may be able to be extracted.  Multiple theories predict the formation of a Mach-cone.  A Mach-cone can be created in hydrodynamics similarly to an airplane creating a sonic boom in air.  In hydrodynamics, a source term is used to generate the energy and/or momentum deposited in the medium as linearized hydrodynamics breaks down near the fast moving parton.  Whether or not a Mach-cone shockwave is created is dependedent of the source term used\cite{mach2, mach4}.  Recently this term has been calculated using pQDC\cite{muller} and from AdS/CFT\cite{ads1}.  A shock wave is then build up through hydrodynamic modes. Figure~\ref{fig:hydro} shows an example of a Mach-cone shock wave from hydrodynamical calculations.  Work is in progress for realistic 3 dimensional hydrodynamic calculations\cite{muller,betz}.

Mach-cones are also be calculated in string theory.  This has been done in AdS/CFT for heavy quarks\cite{ads0} and for quarkonium\cite{ads1} moving through the medium.   For heavy quarks this theory perdicts a cone plus a strong diffusion wake; however, for quarkonium only the cone with no diffusion wake is predicted.  An advantage of this approach is that it does not need a source term.  In the region away from the quark/quarkonia, where linearized hydrodynamics is expected to work, there is good agreement between AdS/CFT and hydrodynamics\cite{ads2}.  There are also Mach-cones predicted in the QCD analog of a QED plasma from plasma physics as longitudinal modes in a strongly interacting colored plasma\cite{mach3}.

The angle of emitted particles is related to the speed of sound a static medium by $c_s/v_{parton}=\cos(\theta_M)$, where $c_s$ is the speed of sound in the medium, $v_{parton}$ is the parton velocity which can be approximated as $c$, and $\theta_M$ is the Mach angle.  The speed of sound of the medium has a temperature dependence\cite{pnjl} so the temperature evolution of the angle needs to be taken into account.  The rapidity distribution and the flow of the dynamic medium, can affect the angle and the shape of the correlation\cite{rr1,rr2}.  

\section{\v{C}erenkov Gluon Radiation}

A particle exceeding the speed of light in the medium can emit gluons in QCD \v{C}erenkov radiation, which would be emitted in a cone around the fast moving parton.  Similar to Mach cone emission the angle of the emitted particles is related to the speed of light in a static medium by $c_{\gamma}/v_{parton}=\cos(\theta_{\check{C}})$ where $c_{\gamma}$ is the speed of light in the medium, and $\theta_{\check{C}}$ is the \v{C}erenkov angle.  The speed of light in the medium has a dependence on the parton velocity.   If the \v{C}erenkov gluons are emitted via resonances, this can give the \v{C}erenkov angle a strong dependence on the emitted particle velocity\cite{cerenkov2}.

\section{Experimental Results}

\begin{figure}[htpb]
\centering
\includegraphics[width=.9\textwidth]{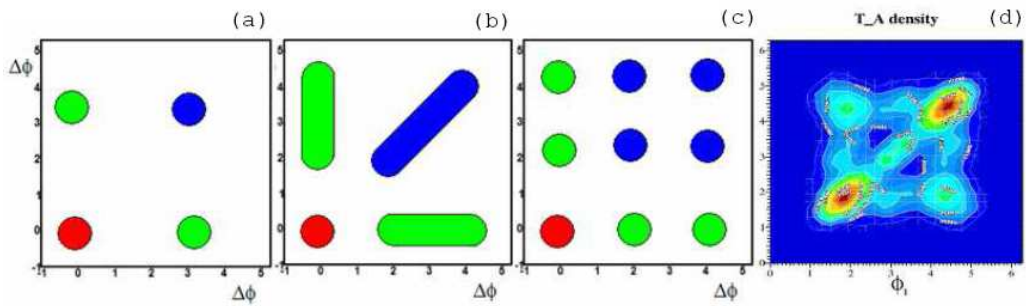}
\caption{Cartoons of 3-particle correlation signals for: (a) back to back jets, (b) deflected jets, (c) conical emission.  (d) Simulation of 3-particle correlations from a cone up in a hydrodynamic expansion\cite{rr2}.}
\label{fig:cartoon}
\end{figure}

Three-particle correlations can yield additional information that can be used to distinguish conical emission from other mechanisms that can describe the 2-particle correlations.  This can be done through the correlation between a trigger particle and two associated particles in $\Delta\phi_1-\Delta\phi_2$ space where $\Delta\phi_i=\phi_{i}-\phi_{Trig}$ is the difference in the azimuthal angles of the associated particle and the trigger particle.  Figure~\ref{fig:cartoon} shows a cartoon example of the 3-particle correlation for different scenarios.  For each case there is a peak at (0,0) where both associated particles are close to the trigger particle.  The peaks at (0,$\pi$) and ($\pi$,0) are from the case where one associated particle is near the trigger and the other is on the away side.  The particles near ($\pi$,$\pi$) are instances where both associated particles are on the away side.  For back-to-back di-jets, just these 4 peaks are present.  If the away side particles are no longer at $180^\circ$ from the trigger due to either path-length dependent energy loss or push from radial flow, then the particles are still columnated.  This would give rise to an away-side peak on the diagonal but displaced from $\pi$.  Over many events this would give a diagonal structure on the away side.  For conical emission, when the cone is projected to the azimuthal plane particles are predominately populated to both sides of $\pi$.  This results in 4 away-side peaks.  The 2 off-diagonal peaks are only present for conical emission and therefore the signature.  Figure~\ref{fig:cartoon}d shows a 3-particle correlation plot of a cone evolved in a hydrodynamic expansion.  This calculation shows how the interaction of the cone with flow can change the signal. 

\subsection{PHENIX Analysis}
\begin{figure}[htbp]
\centering
\includegraphics[width=0.8\textwidth]{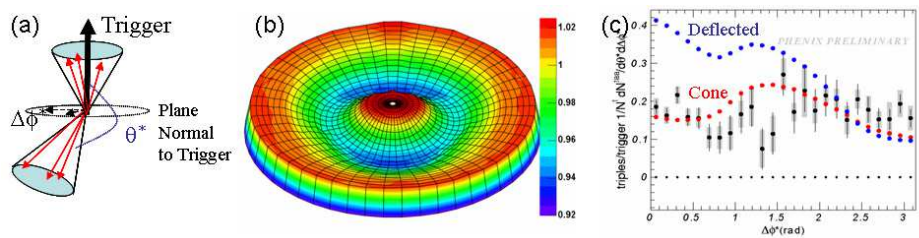}
\caption{Figures for the PHENIX analysis, data plots are for 10-20\% Au+Au collisions.  (a) Diagram of the coordinate system used.  (b) Raw 3-particle correlation. (c)  Projection to the $\Delta\phi^*$ axis of the background subtracted 3-particle correlation.}
\label{fig:phenix}
\end{figure}

The PHENIX collaboration analyzed 3-particle correlations using a $2.5<p_T<4$ GeV/c trigger particle and two $1<p_T<2.5$ GeV/c associated particles\cite{ajitanand}.  This analysis uses a coordinate system where one coordinate is along the direction of the trigger particle momentum.  As illustrated in Fig.~\ref{fig:phenix}a, the angle $\theta^*$ is the angle between the trigger particle and the associated particle.  The other angle $\Delta\phi^*$ is the angle between the two associated in the plane normal to the trigger particle direction.  If the near-side and away-side peaks are back to back in $\eta$ and $\phi$ then this is an excellent coordinate system to view a cone.  However, away-side peak is back to back in azimuth but not in rapidity.  This is because for a collision of two partons they can each carry different fractions of the parents momentum, x, into the collision.  This gives the produced particles an overall momentum along the z-axis.  Figure~\ref{fig:phenix}b shows a raw 3-particle correlation.  The central peak is a near-side peak.  The outer edge are particles $180^\circ$ from the trigger.  Backgrounds from flow and from 2-particle correlations must be subtracted to to observe the jet-like 3-particle correlations.  A $\Delta\phi^*$ projection is shown for the background subtracted 3-particle correlations in Fig.~\ref{fig:phenix}c.  The shapes from their toy model simulations of deflected jets and conical emission are overlaid.  The shape in data is more consistent with the conical emission then the deflected jets.  It should be mentioned however, that the details of the background subtraction and systematics in this analysis are not yet public.

\subsection{STAR Analysis}

\begin{figure}[htpb]
\centering
\includegraphics[width=1.0\textwidth]{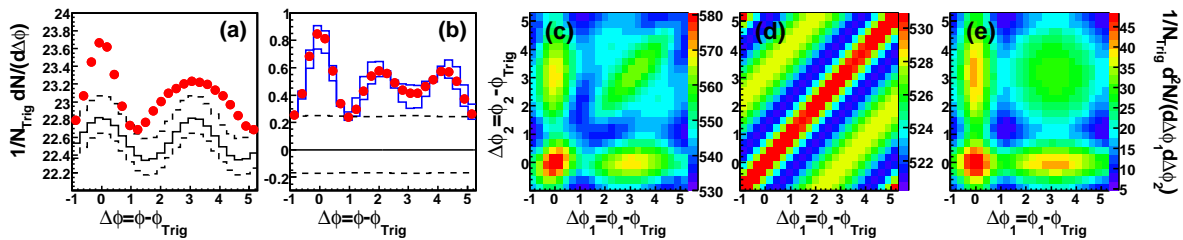}
\caption{(a) Raw 2-particle correlation signal (points), normalized background from mixed events with flow (solid line), uncertainty on the normalization (dashed lines). (b) Background subtracted 2-particle correlation (points), uncertainty on normalization (dashed lines), uncertainty on flow measurements (solid blue lines).  (c) Raw 3-particle correlation signal.  (d) Background from mixing a trigger with associated particles from another events.  (e) Backgrounds from one associated non-flow correlated with the trigger and from flow correlations between the trigger and associated particles.}
\label{fig:bg}
\end{figure}

A 3-particle correlation analysis from STAR uses the assumption the event can be decomposed to particles that are only correlated with the trigger particle via the reaction plane correlation and particles that have other correlations with the trigger particle (such as jet and resonance correlations)\cite{proc1,proc2,proc3,proc4,proc5,writeup,thesis}.  This is done for trigger particles of $3<p_T<4$ GeV/c with pairs of associated particles of $1<p_T<2$ GeV/c.  The raw signal, Fig.~\ref{fig:bg}c, consists of triplets where all are flow correlated, where two are additionally correlated and the third is flow correlated, and where all are additionally correlated.  The signal we are interested in is where all three particles are correlated by more then just the flow correlation so the others must be subtracted.  

To obtain the background where one associated particle is correlated with the trigger particle in we use 2-particle correlations.  Figure~\ref{fig:bg}a shows the raw 2-particle correlations and its background from mixed events with anisotropic flow from $v_2$ and $v_4$ (the second and fourth terms of the Fourier expansion) added from measured values\cite{flow}.  The dashed lines represent the uncertainty on the background normalization.  The 2-particle correlation background is normalized, by factor $a$, such that the final 3-particle correlation signal has zero yield at minimum (ZYAM).  The 2-particle ZYAM assumption is included in the systematics.  Figure~\ref{fig:bg}b shows the background subtracted 2-particle correlation and the uncertainites on normalization and flow. This background is obtained from folding the background subtracted 2-particle correlation with its mixed event background.  

There are also background correlations between the 2 associated particles.  These are obtained from mixing a trigger particle with the associated particles from another event.  Since the associated particles are from the same event all of the correlations that are independent of the trigger particle are preserved. This background is shown in Fig.~\ref{fig:bg}d.  Since the trigger particle and the associated particles also have a correlation with the reaction plane that is broken through event mixing, the flow correlation between the trigger particle and associated particles is added in from measured values for $v_2$ and $v_4$.  These terms are normalized by $a^2b$ where $a$ is the two-particle correlation normalization and $b=[\langle N_{Trig}(N_{Trig}-1)\rangle /\langle N_{Trig} \rangle^2]/[\langle N_{Assoc}(N_{Assoc}-1)\rangle /\langle N_{Assoc} \rangle^2]$ accounts for non-Poisson fluctuations.  The assumption is made that the underlying background deviates from Poisson statistics to a similar degree as the triggered events.  Fig.~\ref{fig:bg}e shows these flow terms along with background where one particle is non-flow correlated with the trigger particles.  The sum of these terms changes little when different $v_2$ and $v_4$ values used.
  
\begin{figure}[htpb]
\centering
\includegraphics[width=1.0\textwidth]{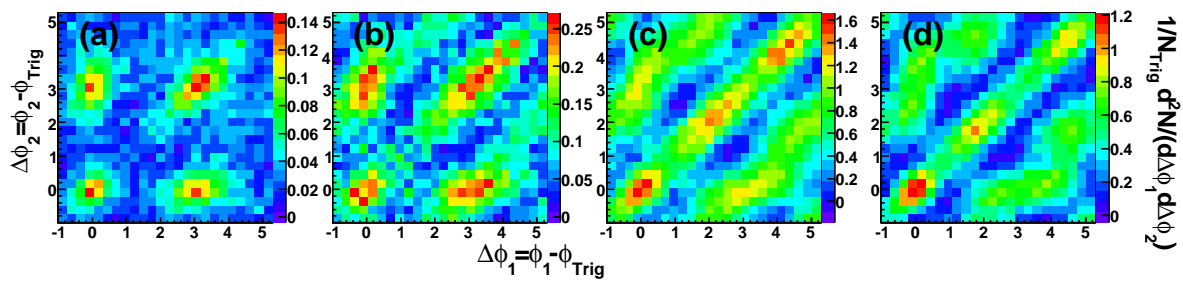}
\caption{Background subtracted 3-particle correlations in (a) d+Au, (b) Au+Au 50-80\%, (c) Au+Au 10-30\%, (d) Au+Au 0-12\% central triggered collisions.}
\label{fig:mine}
\end{figure}

Figure~\ref{fig:mine} shows background subtracted 3-particle correlations in d+Au, and 3 centralities of Au+Au.  The d+Au correlations look similar to the cartoon for back to back jets with a near-side peak at (0,0), away-side peak at ($\pi$,$\pi$), and peaks where one particle is on the near side and the other is on the away side at (0,$\pi$) and ($\pi$,0).  There is, however, some elongation along the diagonal.  This elongation is consistent with $k_T$ broadening, where the near-side and the away-side peaks are not quite back to back in azimuth due to the partons having a small amount of transverse momentum from movement inside the nucleon.  There is additional $k_T$ in d+Au from nuclear $k_T$.  In perpherial Au+Au collisions, additional elongation along the diagonal is seen.  This may be due to some combination of the trigger particle direction differing from the direction of the jet-like correlation axis and deflected jets.  Moving towards more central Au+Au collisions peaks appear on the off-diagonal where peaks are only expected from conical emission.  These peaks are very apparent in the 0-12\% central triggered data set.  

\begin{figure}[htbp]
\centering
\begin{minipage}[t]{0.62\textwidth}
\includegraphics[width=1.0\textwidth]{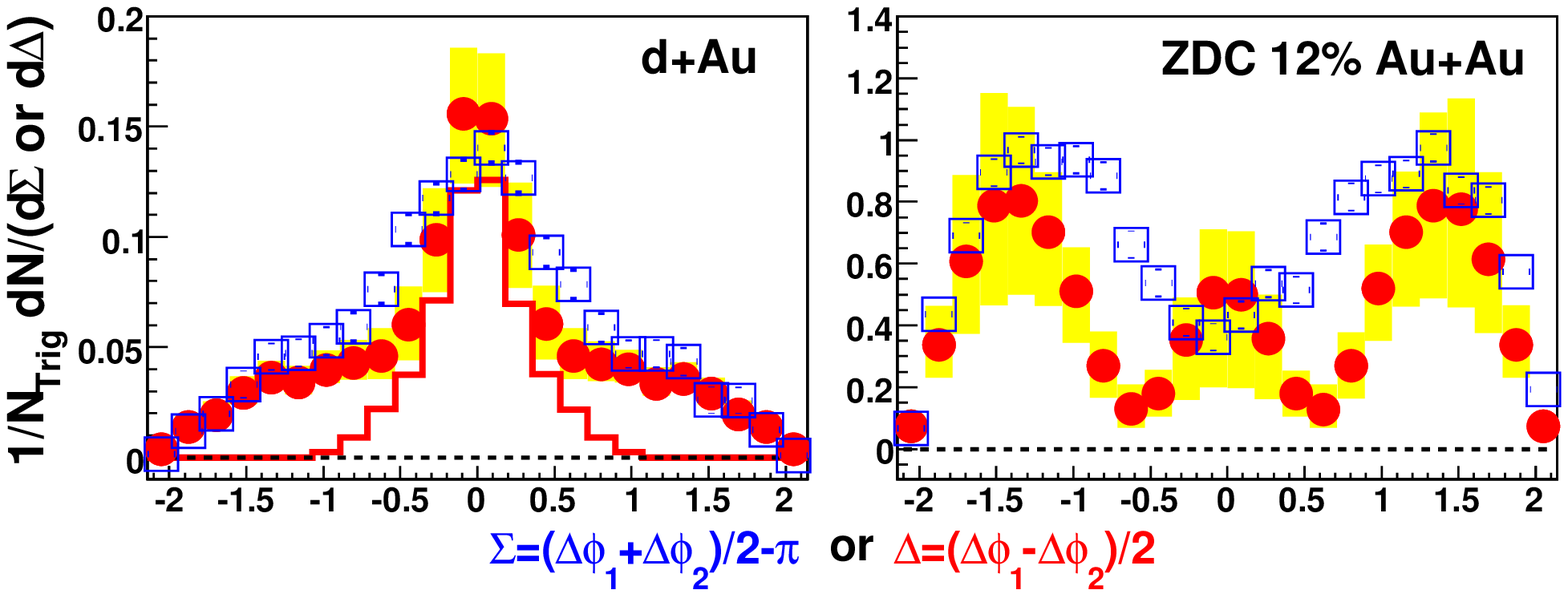}
\end{minipage}
\begin{minipage}[t]{0.37\textwidth}
\includegraphics[width=1.0\textwidth]{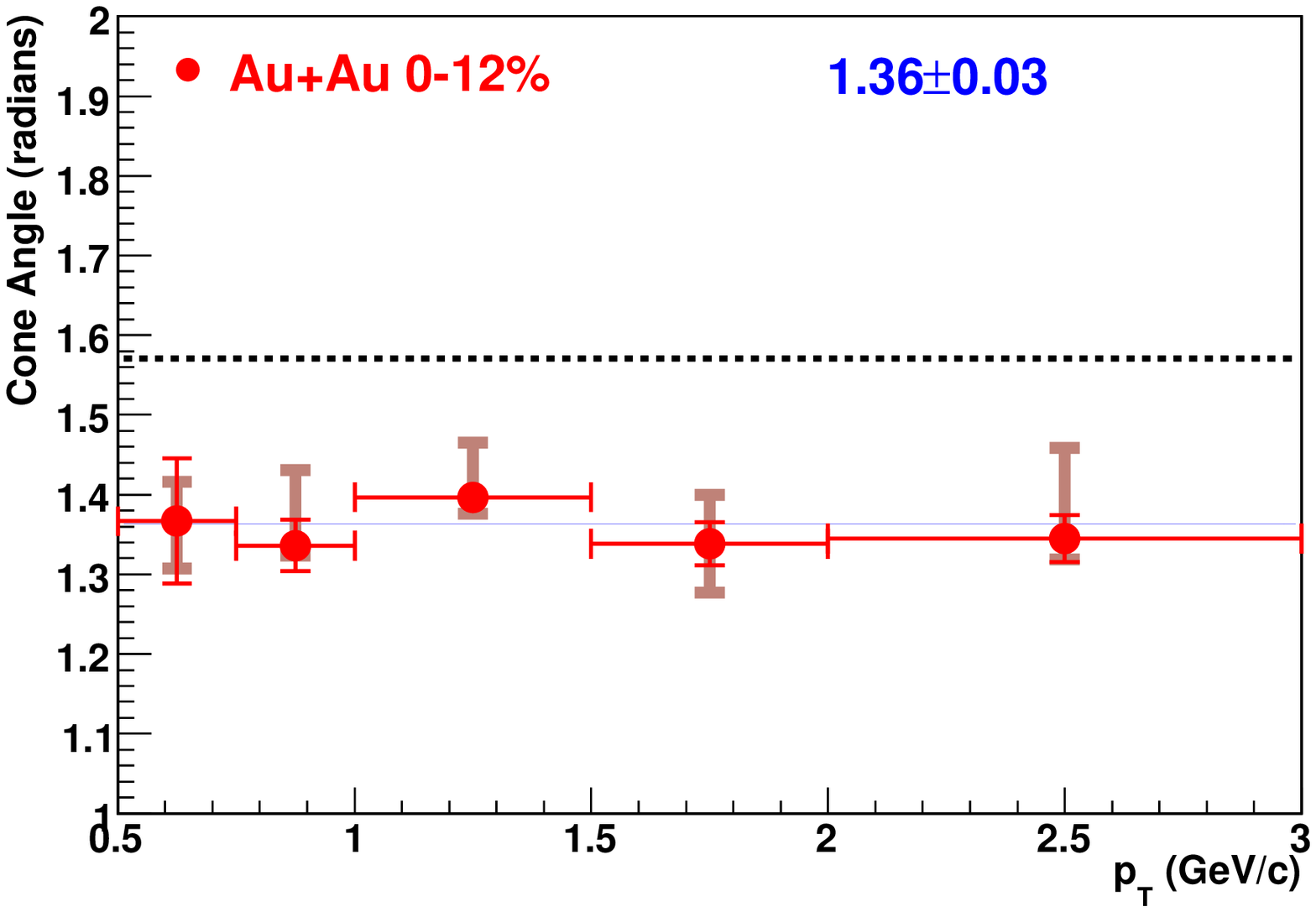}
\end{minipage}
\caption{Left and center: Diagonal projections along $\Sigma$ for $0<\Delta<0.35$ (open squares) and off-diagonal along $\Delta$ for $|\Sigma|<0.35$ (points).  Here $\Sigma=(\Delta\phi_1+\Delta\phi_2)/2-\pi$ and $\Delta=(\Delta\phi_1-\Delta\phi_2)/2$. The shaded bands represent the systematic uncertainites in the off-diagonal projection.  The histogram in d+Au is the near-side off-diagonal projection.  Right: Observed conical emission angle as a function of associated particle $p_T$ in 0-12\% central triggered Au+Au collisions.  Solid bars are statistical error.  Shaded bars are systematic uncertainty.  Dashed line is at $\pi/2$ and solid line is a fit to a constant.  The number 1.36$\pm$0.03 is the value from the fit.}
\label{fig:proj}
\end{figure}

To study the 3-particle correlation signal in detail, projections are taken on the away side along the diagonal and off-diagonal.  The projections for d+Au and central Au+Au collisions are shown in Fig.~\ref{fig:proj}.  The diagonal projection of the near-side in d+Au is shown for comparison.  In d+Au, all of the projections have a single peak.  The peak in the diagonal projection is broader then the off-diagonal likely due to $k_T$ broadening.  In central Au+Au, the diagonal projection displays a double peaked structure similar to what is seen on the away side in 2-particle correlations.  The off-diagonal projection shows 3 peaks.  A central peak and 2 symmetric side peaks.  The side peaks are the signature for conical emission. These peaks can be fit with a Gaussian for the central peak and another for the side peaks.  The position of this Gaussian can be used to extract the conical emission angle.  For 0-12\% central Au+Au the observed angle is $1.38\pm0.02$(stat.)$\pm0.06$(sys.).  This angle can also be studied as a function of associated particle $p_T$.  This is important because calculations for \v{C}erenkov gluon radiation emitted via resonances predict a sharply decreasing angle with associated particle momentum.  Figure~\ref{fig:proj}(right) shows the associated particle $p_T$ dependence of the extracted angle.  No significant dependence is observed which is consistent with the expectation of Mach-cone shock-wave and inconsistent with the expectation from \v{C}erenkov gluon radiation via resonances.  

\subsection{CERES Analysis}

\begin{figure}[htpb]
\centering
\includegraphics[width=1.0\textwidth]{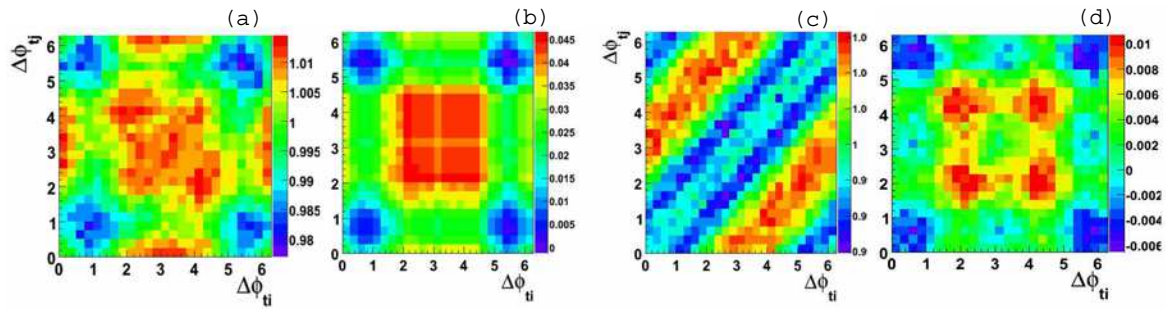}
\caption{Three-particle correlation results from CERES.  (a) Raw signal.  (b) Background where one of the associated is non-flow correlated with the trigger particle.  (c) Background containing correlations between the associated particles that are independent of the trigger particle.  (e) Background subtracted 3-particle correlations.}
\label{fig:sps}
\end{figure}

The 3-particle correlation analysis from CERES uses the same method as the STAR analysis\cite{stephen}.  The analysis uses a trigger particle of $2.5<p_{T}<4$ GeV/c with pairs of associated particles of $1<p_T<2.5$ GeV/c.  The raw signal is shown in Fig.~\ref{fig:sps}a.  Backgrounds are removed to account for one of the particles correlated with the trigger (Fig.~\ref{fig:sps}b) and for correlations between the associated particles that are independent of the trigger particle (Fig.~\ref{fig:sps}c).  The background from the $v_2$ correlation between the trigger and associated particles is removed.  Figure~\ref{fig:sps}d shows the background subtracted 3-particle correlations signal.  There are clear off-diagonal peaks indicating the presence conical emission at the SPS also.  Systematic uncertainties have not been fully evaluated for this analysis presently.

\section{Summay}

There has been renewed interest in conical emission due to the observed modification of the away-side structure in 2-particle jet-like correlations.  This modification can be explained through many physics mechanisms.  Two of the mechanisms are Mach-cone shock waves and \v{C}erenkov gluon radiation which have conical emission.  Conical emission will have distinctive features in azimuthal 3-particle correlations that have been exploited by experiments to identify the physics mechanisms underlying the away-side modification in jet-like correlations. There are results from two experiments at RHIC and one from SPS.  The PHENIX experiment shows results that are more consistent with their toy model simulation of conical emission then of deflected jets.  However further systematic checks are necessary.  The STAR results show significant conical emission peaks in central Au+Au collisions at about 1.38 radians.  The independence of this angle on associated particle $p_T$ is consistent with Mach-cone shock waves and inconsistent with \v{C}erenkov gluon radiation via resonances.  The CERES experiment shows conical emission peaks that are consistent with the STAR results.  There has been much progress in theoretical work in Mach-cone shock wave generation and propogation in relativistic heavy ion collision.  Further study is required in order to make a qualitative statement about the equation of state from the experimental measurements.

\end{document}